\documentclass[twocolumn,aps,twoside,preprintnumbers]{revtex4}
\usepackage{amsmath,amssymb,amsthm}
\usepackage{latexsym,graphicx,color,bbm,subfigure}

\newcommand{\NF}{N_{\rm f}}

\newcommand{\beq}{\begin{eqnarray}}
\newcommand{\eeq}{\end{eqnarray}}
\newcommand{\non}{\nonumber\\}
\newcommand{\D}{\mathcal{D}}

\newcommand{\Tr}{{\rm Tr}}
\newcommand{\diag}{{\rm diag}}

\begin{document}

\title{Low-energy $U(1)\times USp(2M)$ gauge theory
from simple high-energy gauge group}
\author{Sven Bjarke Gudnason}
\email{gudnason(at)df.unipi.it}
\author{Kenichi Konishi}
\email{konishi(at)df.unipi.it}
\affiliation{Department of Physics, E.~Fermi, University of Pisa, \\
INFN, Sezione di Pisa,\\
Largo Bruno Pontecorvo, 3, Ed.~C, 56127 Pisa, Italy}
\begin{abstract}
\preprint{IFUP-TH-2010-06}

We give an explicit example of the embedding of a near BPS low-energy 
$(U(1)\times USp(2M))/\mathbb{Z}_2$ gauge theory into a high-energy
theory with a simple gauge group and adjoint matter content. This
system possesses degenerate monopoles arising from the high-energy
symmetry breaking as well as non-Abelian vortices due to the symmetry
breaking at low energies. These solitons of different codimensions are
related by the exact homotopy sequences. 
\end{abstract}
\maketitle

\section{Introduction}

Topological solitons are important in many areas of physics, ranging 
from high-energy (elementary particle) physics,  condensed matter
physics and string theory to cosmology. In this letter, we shall focus
on a system possessing non-Abelian vortices and monopoles: a
supersymmetric gauge theory with $G=USp(2M)$ gauge group, which is
broken to $H=U(1)\times USp(2M-2)$ by the vacuum expectation value
(VEV) of an adjoint scalar field. This breaking gives rise to regular
non-Abelian 't Hooft-Polyakov monopoles. According to
Goddard-Nuyts-Olive-Weinberg 
\cite{Goddard:1976qe,Bais:1978fw,Weinberg:1979zt,Weinberg:1982ev}, 
the non-Abelian monopoles transform according to the dual group of $H$
which in this case is $\tilde{H}= U(1)\times SO(2M-1)$. Several
difficulties in the na\"{i}ve idea of non-Abelian monopoles have been
known for some time, i.e.~the global $H$ group suffering from a
topological obstruction and non-normalizable zero-modes do not allow 
the standard quantization and construction of the $H$ multiplets of 
monopoles
\cite{Abouelsaood:1982dz,Nelson:1983bu,Balachandran:1982gt,Horvathy:1984yg,Horvathy:1985bp}. 
These problems arise in the Coulomb phase of the theory. 

As was done in a series of investigations
\cite{Auzzi:2003fs,Auzzi:2003em} for $SU(N)$ gauge theories, we take
one step further here and break the remaining gauge symmetry
completely at a much lower mass scale. This can be realized by the
introduction of an ${\cal N}=2$ breaking term in the superpotential,
giving rise to an effective Fayet-Iliopoulos term. In systems with
such a hierarchical gauge symmetry breaking, the homotopy group-maps
relate the regular monopoles to the non-Abelian vortices arising at
low energies, allowing for a better understanding of the concept of 
the non-Abelian monopole itself.  Also, this kind of system provides
a (dual) model of a non-Abelian color-confining superconductor,
further motivating its study.

Besides the cases of $SU(N)$ gauge theories extensively studied in
the last several years, this type of analysis has so far been made
only in the case of $SO(N)$ gauge theories \cite{Ferretti:2007rp},
i.e.~with a hierarchical breaking, 
$SO(N) \to U(1)\times SO(N-2)\to {\mathbbm 1}$. In the $SO(N)$ systems 
the adjoint matter in the high-energy system yields at low energies
exactly the right matter content -- a system with light fundamental
matter, all charged with respect to a common $U(1)$ factor. 

As gauge systems with hierarchical symmetry breaking 
$G \to H \to {\mathbbm 1}$ and a color-flavor locked symmetry
$H_{C+F}$, have been constructed to date only for the $SU(N)$ and
$SO(N)$ gauge groups \cite{Auzzi:2003fs,Auzzi:2003em,Ferretti:2007rp}, 
one might wonder to which extent our idea of defining non-Abelian
monopoles through better-understood non-Abelian {\it vortices}
is general. 
The central aim -- and the result -- of the present note is to
construct explicitly an analogous system with the unitary symplectic
gauge group, strengthening further our belief that this kind of
approach is of quite a general validity. 

Among the many remarkable developments which followed the discovery of
genuine non-Abelian vortices (vortices with continuous non-Abelian
moduli) in Refs.~\cite{Hanany:2003hp,Auzzi:2003fs} is the moduli
matrix formalism \cite{Isozumi:2004jc,Eto:2005yh} (see review
\cite{Eto:2006pg}), first constructed 
for domain walls. This formalism made it possible to uncover the full
moduli space of these non-Abelian vortices, first in the 
$U(N)\sim (U(1)\times SU(N))/\mathbb{Z}_N$ theories and subsequently
in models with generic gauge groups \cite{Eto:2008yi}. Finally, in
Ref.~\cite{Eto:2009bg} an in-depth study of the non-Abelian vortices
including the cases of the $(U(1)\times USp(2M))/\mathbb{Z}_2$ gauge
group, has been carried out. 

The system considered in this note reduces  at low energies, as we
shall show, to the 
$(U(1)\times USp(2M))/\mathbb{Z}_2$ models investigated in
Ref.~\cite{Eto:2009bg}; the properties of the vortex moduli space
found there then give detailed exact information about the massive
non-Abelian monopoles. 

\section{$USp(2M)$ theory with matter in the fundamental
 representation} 

Let us first briefly review the superpotential for $\NF$ fundamental 
hypermultiplets in the $USp(2M)$ gauge theory with $\mathcal{N}=2$
extended supersymmetry in $3+1$ dimensions  
\beq \sqrt{2}\sum_{i=1}^{\NF}\tilde{q}_a^i\Phi^{ab}q_b^i \ , \eeq
where $i$ denotes the flavor index and $a,b=1,\ldots,2M$ denote the
color indices.
Due to the pseudo-real nature of $USp$ matter fields, we can by a
change of basis 
\beq q^i = \frac{1}{\sqrt{2}}\left(Q^i+iQ^{\NF+i}\right) \ , \ \
\tilde{q}^i = \frac{1}{\sqrt{2}}\left(Q^i-iQ^{\NF+i}\right) \ , 
\label{eq:basischange}
\eeq
write the superpotential as
\beq \frac{1}{\sqrt{2}}\sum_{i=1}^{2\NF} Q_a^i\Phi^{ab}Q_b^i \ , \eeq
where we have used the fact that $\Phi^{ab}=\Phi^{ba}$ is symmetric
and we use a notation where the color indices are raised and lowered
with the invariant rank-two tensor of $USp(2M)$
\beq J^{\rm T} = - J \ , \ \ J^\dag J = \mathbf{1}_{2M} \ , \eeq
which we choose to be the skew-diagonal matrix as usual.
The (global) flavor symmetry which the theory at hand possesses is
$O(2\NF)$.  
The mass term is
\beq \sum_{i,j=1}^{2\NF}\frac{m_{ij}}{2}Q_a^iJ^{ab}Q_b^j \ , \eeq
where $m_{ij}=\hat{m}_iJ_{ij}$ is anti-symmetric. The flavor symmetry
is now $O(2\NF)\cap USp(2\NF)\sim U(\NF)$. 

\section{$USp(2M)$ theory with matter in the adjoint representation }

To construct a system with a hierarchical gauge symmetry breaking as
explained in the Introduction we use the matter fields (squarks) in the
adjoint representation rather than in the fundamental
representation. As in the previous case we start with the matter
fields in the basis 
\beq \sqrt{2}\sum_{i=1}^{\NF}\Tr\left\{
\tilde{q}^i\left[\Phi,q^i\right]\right\} \ , \eeq
while by the change of basis (\ref{eq:basischange}) we obtain
\begin{align}
\mathcal{W}_{\rm Adj,Yukawa} &=
\frac{i}{\sqrt{2}}
\sum_{i,j=1}^{2\NF}J_{ij}\Tr\left\{Q^i\left[\Phi,Q^j\right]\right\}
\non
&=i\sqrt{2}\sum_{i,j=1}^{2\NF}J_{ij}\Tr\left\{Q^i\Phi Q^j\right\}
\ ,
\end{align} 
with $J^{\rm T}=-J, \ J^\dag J = \mathbf{1}_{2\NF}$ being the rank-two
invariant tensor of 
$USp(2\NF)$ \footnote{Notice that the flavor symmetry of this system is
different with respect to the usual $O(2\NF)$ as possessed by the
system with fundamental matter multiplets. }, whereas
the mass term is now 
\begin{align}
\sum_{i,j=1}^{2\NF}\frac{m_{ij}}{2}
\Tr\left\{Q^i Q^j\right\} \ , 
\end{align}
and needs to be symmetric in order not to vanish. We shall choose
$m_{ij}=\hat{m}_i \,\tilde{J}_{ij}$, where $\tilde{J}$ is the
symmetric invariant tensor of $SO(2\NF)$
\beq \tilde{J}^{\rm T} = \tilde{J} \ , \ \ \tilde{J}^\dag \tilde{J} =
\mathbf{1}_{2\NF} \ , \eeq 
where we again use the skew-diagonal basis. The global flavor symmetry
of our system is thus $USp(2\NF)\cap O(2\NF)\sim U(\NF)$.

\section{$\mathcal{N}=1$ deformation}

Finally, we will add a soft supersymmetry breaking term as
$\mu\,\Tr\,\Phi^2$ to the adjoint theory and hence we have the
superpotential  
\begin{align}
\mathcal{W}_{\rm Adj} = &\ 
i\sqrt{2}
\sum_{i,j=1}^{2\NF}J_{ij}\Tr\left\{Q^i\Phi Q^j\right\} 
+\!\sum_{i,j=1}^{2\NF}\frac{m_{ij}}{2}
\Tr\left\{Q^i Q^j\right\} \non
&+\frac{\mu}{2}\Tr\left\{\Phi^2\right\} \ ,
\end{align}
which gives rise to the following vacuum equations
\begin{align}
J_{ij}\left[Q^j,\Phi\right] + \frac{i}{\sqrt{2}}m_{ij}Q^j &= 0 \ , 
\ \ i=1,\ldots,2\NF \ , \\
J_{ij}\left[Q^i,Q^j\right] + i\sqrt{2}\mu\Phi &= 0 \ ,
\end{align}
(repeated indices are summed over) together with the $D$-term
conditions. 

First a word on what we expect. From group theory we know that the
adjoint representation of $USp(2M)$ splits as 
\cite{Slansky:1981yr}
\begin{align}
USp(2M) &\supset SU(2)\times USp(2M-2) \ , \\
{\rm Adj} &= ({\rm Adj},\mathbbm{1}) + (\mathbbm{1},{\rm Adj})
+(\square,\square) \ , \nonumber
\end{align}
($M>1$). 
Actually, we are interested only in the $U(1)$ subgroup of $SU(2)$ so
the relevant decomposition reads 
\begin{align}
USp(2M) &\supset U(1)\times USp(2M-2) \ ,   \label{eq:fund} \\
{\rm Adj} &=  3\, (0, {\mathbbm 1})  +   (0,{\rm Adj})
+(1,\square) + (-1,\square) \ . \nonumber
\end{align}
We require the system to be such that only the fields in the
fundamental representation in the low-energy $USp(2M-2)$ remain light,
other fields with no $U(1)$ charges all becoming massive, with a mass
of the order $\mathcal{O}(m)$. Furthermore, only one set of
fundamentals will remain light, either the one with positive $U(1)$
charge or the one with negative charge in Eq.~(\ref{eq:fund}). 

We choose the VEV of $\Phi$ as
\beq \langle\Phi\rangle = \epsilon \ 
\diag(m,\underbrace{0,\ldots,0}_{M-1},-m,\underbrace{0,\ldots,0}_{M-1})
\equiv \epsilon\, \Phi_0 \ , \ \eeq
and the mass parameters as 
\beq m_{ij}= -i\sqrt{2}\,m \, \tilde{J}_{ij} \ , \ \ 
\mu = -i\sqrt{2} \, \nu \ , \label{massterms}   \eeq
where again $\tilde{J}=\tilde{J}^{\rm T}$ is the invariant tensor of
$SO(2\NF)$. 
In order to have a separation of scales in the hierarchical gauge
symmetry breaking, we take $m\gg \nu$.
$\epsilon=\pm$ is the sign that will select which
fundamental fields will become light, with positive or
negative $U(1)$ charge, respectively.  
Accordingly, we make an Ansatz $Q^{\NF+i}=(Q^i)^\dag$, which solves
the $D$-flatness conditions. 
This Ansatz together with the masses taken as in Eq.~(\ref{massterms})
reduces the vacuum equations to  
\begin{align}
\left[\Phi_0,Q^i\right] + \epsilon \, m \, Q^i &= 0 \ , \ \ i=1,\ldots,\NF\ , 
\label{eq:vac1}\\
\sum_{i=1}^{\NF}\left[Q^i, Q^{i\dag}\right] + \epsilon\, \nu\, \Phi_0 &= 0 \ .
\label{eq:vac2}
\end{align}
The light fields are then seen to correspond to the non-trivial
eigenvectors of $[ \Phi_0, \cdot\,]$ with eigenvalue $-\epsilon\, m$ and
they in turn condense by Eq.~(\ref{eq:vac2}). 
Without loss of generality, we can choose the light 
fields to be the ones with positive $U(1)$ charge and set
$\epsilon:=+$. 
Such eigenvectors are found to be $h^{i}(x)$ where  
\begin{align}
Q^i = Q_a^i \, t^a = h_{\alpha}^{i} \, K^\alpha +
h_{M-1+\alpha}^{i}\,  L^\alpha \ ,  
\end{align}
where $i=1,\ldots,\NF$ is the flavor index and $a=1,\ldots,M(2M+1)$
is the adjoint color index and finally $\alpha=1,\ldots,M-1$ is half of
the fundamental color index for $USp(2M-2)$. The matrices
$K,L\in\mathfrak{usp}(2M)^{\mathbb{C}}$ are 
\begin{align}
{(K^\alpha)_i}^j &= \frac{1}{2}
\left({\delta_{1+\alpha,i}}   \delta^{1,j} - 
{\delta_{M+1,i}} \delta^{M+1+\alpha,j}\right) \ ,
\\
{(L^\alpha)_i}^j &= \frac{1}{2}
\left({\delta_{M+1+\alpha,i}} \delta^{1,j} + 
{\delta_{M+1,i}} \delta^{1+\alpha,j}\right) \ .
\end{align}
If we instead wanted the fundamental fields with negative $U(1)$
charge to be the light fields, we should set $\epsilon:=-$ and the
eigenvectors would be
\begin{align}
Q^i = Q_a^i \, t^a = h_{\alpha}^{i} \, \left(K^\alpha\right)^{\rm T} +
h_{M-1+\alpha}^{i}\,  \left(L^\alpha\right)^{\rm T} \ ,   
\end{align}
see Appendix for details. 

Calculating now explicitly the commutator, Eq.~(\ref{eq:vac2}) gives
rise to the $D$-flatness conditions of the $U(1)\times USp(2M-2)$
low-energy 
theory with fundamental matter content. 
Let us make the following definition 
\beq h^i  = \begin{pmatrix} h^i_{\alpha}
 \\   h^i_{M-1+\alpha}   \end{pmatrix}    \equiv \begin{pmatrix}
 k^i_{\alpha} \\ \ell^i_{\alpha} \end{pmatrix} \ , \eeq 
with $k,\ell$ being $(M-1)$-vectors of color and $i$ is the flavor
index. Then, independently of the choice of the sign $\epsilon$,
($4\times$) Eq.~(\ref{eq:vac2}) reads 
\begin{align}
\begin{pmatrix}
-h^{i\dag}h^i + 4\nu m & 0 &
0 & 0\\
0 & \mathbf{A} & 0 & \mathbf{B}^\dag \\
0 & 0 & h^{i\dag}h^i - 4\nu m & 0\\
0 & \mathbf{B} & 0 & -\mathbf{A}^{\rm T} 
\end{pmatrix} = 0 \ , 
\label{eq:vacfinal1}
\end{align}
from which the Abelian $D$-term
constraint (in the low-energy $\mathcal{N}=1$ theory) is easily read
off. Now for the non-Abelian part, we find the form of the matrices 
${\bf A}, {\bf B}$:
\begin{align}
\mathbf{A} \equiv k^i k^{i\dag} -
\left(\ell^i \ell^{i\dag}\right)^{\rm T} \ , \ \ 
\mathbf{B} \equiv \ell^i k^{i\dag} + 
\left(\ell^i k^{i\dag}\right)^{\rm T} \ ,
\end{align}
where ${\bf B}^{\rm T}={\bf B}$ is manifest. 
Using that
\beq h^i h^{i\dag} = 
\begin{pmatrix}
k^i k^{i\dag} & k^i \ell^{i\dag} \\
\ell^i k^{i\dag} & \ell^i \ell^{i\dag} 
\end{pmatrix} \ , \label{eq:hhdag_explicit}
\eeq
together with the explicit form of the generators
\beq
t^n = 
\begin{pmatrix}
\alpha & \beta_{S} \\
\beta_{S}^\dag & -\alpha^{\rm T} 
\end{pmatrix} \ , \ \ 
\alpha^\dag = \alpha \ , \ \ 
\beta_{S}^{\rm T} = \beta_{S} \ , 
\eeq
we obtain
\begin{align}
0 &= \Tr\left\{h^i h^{i\dag}t^n\right\} \non
&= \Tr\left\{{\bf A}\alpha\right\} +
\frac{1}{2}\Tr\left\{{\bf B}\beta_{S}\right\} +
\frac{1}{2}\Tr\left\{{\bf B}^\dag\beta_{S}^\dag\right\} \ ,
\label{eq:Dterm2}
\end{align}
for \emph{all} $\alpha,\beta_{S}$, which forces ${\bf A}={\bf B}=0$,
where we have used the fact that ${\bf B}$ is symmetric. 
Now as a check, we can count the number of constraints of 
${\bf A}={\bf B}=0$ yielding $M'(2M'+1)$ with $M'\equiv M-1$, which
indeed  
coincides with the number of constraints in Eq.~(\ref{eq:Dterm2}). 
Hence, using a color-flavor matrix notation 
${(h h^\dag)_{\alpha}}^{\alpha'} = {h_\alpha}^i {(h^\dag)_i}^{\alpha'}
= h^i h^{i\dag}$ we can write the
Eqs.~(\ref{eq:vac1})-(\ref{eq:vac2}) as 
\begin{align}
\Tr\left\{h h^\dag\right\} &= 4 \nu m \ , \label{eq:Dflat1} \\
\Tr\left\{h h^\dag t^n\right\} &= 0 \ , \label{eq:Dflat2}
\end{align}
which are the $D$-term conditions appropriate for constructing
non-Abelian BPS vortices and $t^n\in\mathfrak{usp}(2M-2)$ and
$n=1,\ldots,(M-1)(2(M-1)+1)$ and specifically for the fundamental
representation, as we intended. 
These vortices have already been studied in the low-energy theory in
Ref.~\cite{Eto:2009bg}. 
A comment in store is to emphasize the importance of identifying the
``light mass'' degrees of freedom in the symmetry breaking. 

In order to have a vacuum that breaks completely the local gauge
symmetry, allowing at the same time for an intact global color-flavor
symmetry, we shall choose the number of flavor multiplets to be
$\NF=2M-2$. Thus $h$ is a square matrix with the following VEV
\beq \langle h\rangle = 
\frac{\sqrt{\xi}}{\sqrt[4]{M-1}}\mathbf{1}_{2M-2} \ . \eeq

For completeness, let us write down the low-energy effective action
for the light fundamental fields
\begin{align}
\mathcal{L} =\;& 
-\frac{1}{4g^2}F_{\mu\nu}^n F^{n\mu\nu}
-\frac{1}{4e^2}F_{\mu\nu}^0 F^{0\mu\nu}
+\Tr\left(\D_\mu h\right)\left(\D^\mu h\right)^\dag \non &
-\frac{e^2}{2}\left|\Tr\left(h h^\dag t^0\right) - \xi\right|^2
-\frac{g^2}{2}\left|\Tr\left(h h^\dag t^n\right)\right|^2 \ ,
\end{align}
where we have rescaled the fields $h\to \sqrt{2}gh$ and 
$\xi\equiv \nu m \to e\,\xi/(\sqrt{2M-2})$ and defined the $U(1)$
generator 
\beq t^0 \equiv \frac{\mathbf{1}_{2M-2}}{2\sqrt{M-1}} \ , \eeq 
and finally the index $n=1,\ldots,(M-1)(2(M-1)+1)$. 
Due to different renormalization effects of the subgroups after
the gauge symmetry breaking, we use $e$ to denote the coupling for
$U(1)$ and $g$ for $USp(2M-2)$. Note that we have neglected higher
order terms in $\nu/m$ which will give rise to non-BPS terms in the
low-energy action for vortices, hence as already mentioned it is a
near-BPS system. 
 
As a final remark, let us note that in the strictly BPS limit our
low-energy system would have a large vortex-moduli space including the 
so-called semi-local vortices \cite{Eto:2009bg}. The latter do not
have a definite transverse size, and would not confine the monopole at 
their ends.  
However, our system (with $m \gg \nu$) is almost, but not exactly,
BPS. When small non-BPS corrections arising from the high-energy gauge
symmetry breaking are taken into account, we expect the vortex moduli,
which are not related to the exact global symmetry of the system, to
disappear.   
This has been explicitly shown \cite{Auzzi:2008wm} in the case of the
vortex moduli in the $SU(N+1)$ theory with $N_{f}> N$,
spontaneously broken at two scales,  
$SU(N+1) \to U(N) \to {\mathbbm 1}.$

\section{Conclusion}

Our system is characterized by the hierarchical gauge symmetry
breaking  
\beq G \ \mathop{\longrightarrow}^{m}\ H \ 
\mathop{\longrightarrow}^{2\sqrt{\nu m}}\ \mathbbm{1} \;.  \eeq
As all the fields in the underlying theory are in the adjoint
representation, we actually have $G=USp(2M)/\mathbb{Z}_2 $.  The
(light) matter content of the low-energy theory shows also that 
$H=\left(U(1)\times USp(2M-2)\right)/\mathbb{Z}_2$.  Since 
$\pi_1(G)=\mathbb{Z}_2$, the exact homotopy sequence tells us that 
\beq \pi_2\left(G/H\right) \sim \pi_1\left(H\right)/ \mathbb{Z}_2    
\;:  \eeq
the regular monopoles arising at the high-mass scale breaking 
are confined by the doubly-wound vortices of the low-energy theory.  
The results of Ref.~\cite{Eto:2009bg}, which hold in a vacuum with the
color-flavor locked phase, indicate that the minimal winding vortices
of the low-energy $U(1)\times USp(2M-2)$ system, which are stable in
the full theory as $\pi_1(G)=\mathbb{Z}_2$, appear classified
according to the {\it spinor representation} of a dual (color-flavor)
$SO(2M-1)$ symmetry group. The regular monopoles of our system,
associated with the doubly-wound vortices, are then predicted 
to transform according to various representations including the 
{\it vector representation} of the $SO(2M-1)$ group, reminiscent of
the GNO duality. These group-theoretic features of our vortex-monopole
complex are under a careful scrutiny at present, and will be presented
elsewhere.

\section*{Acknowledgments} 

We thank Minoru Eto, Muneto Nitta and  Keisuke Ohashi for useful discussions.

\appendix

\section{}
It is also possible, though more elaborate to use the real algebra
$\mathfrak{usp}(2M)$ instead of the complexified algebra
$\mathfrak{usp}(2M)^{\mathbb{C}}$ as we have utilized in the
calculation. However, it requires to change the basis. Using the
definitions we already have made, we can write
\begin{align}
Q_a^i t^a &= 
H_\alpha^i \kappa^\alpha + \tilde{H}_\alpha^i \tilde{\kappa}^\alpha
+ &&H_{M-1+\alpha}^i \lambda^\alpha + \tilde{H}_{M-1+\alpha}^i
\tilde{\lambda}^{\alpha} \ , \nonumber
\end{align}
with
\begin{align}
\kappa^\alpha &\equiv 
 K^\alpha + \left(K^\alpha\right)^{\rm T} \ , \ \ 
&\tilde{\kappa}^{\alpha} &\equiv 
 i K^\alpha - i\left(K^\alpha\right)^{\rm T} \ , \\  
\lambda^\alpha &\equiv 
 L^\alpha + \left(L^\alpha\right)^{\rm T} \ , \ \
&\tilde{\lambda}^{\alpha} &\equiv 
 i L^\alpha - i\left(L^\alpha\right)^{\rm T} \ , 
\end{align}
where
$\kappa,\tilde{\kappa},\lambda,\tilde{\lambda}\in\mathfrak{usp}(2M)$. 
Now to obtain the eigenvectors
in this basis, we find the following linear combination
\beq h_\epsilon^i = 
\frac{1}{\sqrt{2}}\left(H^i + \epsilon i\tilde{H}^i\right) \ . \eeq
We recognize $h_\epsilon^i$ as $\sqrt{2}$ times the eigenvectors found
in the text and $\epsilon$ again selects the $U(1)$ charge.
Thus it is an advantage to work directly with the complexified
algebra.


\begin{thebibliography}{99}

\bibitem{Goddard:1976qe}
P.~Goddard, J.~Nuyts and D.~I.~Olive,
Nucl.\ Phys.\  B {\bf 125}, 1 (1977).
\bibitem{Bais:1978fw}
F.~A.~Bais,
Phys.\ Rev.\  D {\bf 18}, 1206 (1978).
\bibitem{Weinberg:1979zt}
E.~J.~Weinberg,
 ``Fundamental Monopoles And Multi-Monopole Solutions For Arbitrary Simple
Nucl.\ Phys.\  B {\bf 167}, 500 (1980).
\bibitem{Weinberg:1982ev}
E.~J.~Weinberg,
Nucl.\ Phys.\  B {\bf 203}, 445 (1982).
\bibitem{Abouelsaood:1982dz}
A.~Abouelsaood,
Nucl.\ Phys.\  B {\bf 226}, 309 (1983).
\bibitem{Nelson:1983bu}
P.~C.~Nelson and A.~Manohar,
Phys.\ Rev.\ Lett.\  {\bf 50}, 943 (1983).
\bibitem{Balachandran:1982gt}
A.~P.~Balachandran, G.~Marmo, N.~Mukunda, J.~S.~Nilsson, E.~C.~G.~Sudarshan and F.~Zaccaria,
Phys.\ Rev.\ Lett.\  {\bf 50}, 1553 (1983).
\bibitem{Horvathy:1984yg}
P.~A.~Horvathy and J.~H.~Rawnsley,
Phys.\ Rev.\  D {\bf 32}, 968 (1985).
\bibitem{Horvathy:1985bp}
P.~A.~Horvathy and J.~H.~Rawnsley,
J.\ Math.\ Phys.\  {\bf 27}, 982 (1986).
\bibitem{Auzzi:2003fs}
R.~Auzzi, S.~Bolognesi, J.~Evslin, K.~Konishi and A.~Yung,
Nucl.\ Phys.\  B {\bf 673}, 187 (2003)
[arXiv:hep-th/0307287].
\bibitem{Auzzi:2003em}
R.~Auzzi, S.~Bolognesi, J.~Evslin and K.~Konishi,
Nucl.\ Phys.\  B {\bf 686}, 119 (2004)
[arXiv:hep-th/0312233].
\bibitem{Ferretti:2007rp}
L.~Ferretti, S.~B.~Gudnason and K.~Konishi,
Nucl.\ Phys.\  B {\bf 789}, 84 (2008)
[arXiv:0706.3854 [hep-th]].
\bibitem{Hanany:2003hp}
A.~Hanany and D.~Tong,
JHEP {\bf 0307}, 037 (2003)
[arXiv:hep-th/0306150].
\bibitem{Tong:2005un}
  D.~Tong,
  arXiv:hep-th/0509216.
\bibitem{Shifman:2007ce}
  M.~Shifman and A.~Yung,
  Rev.\ Mod.\ Phys.\  {\bf 79}, 1139 (2007)
  [arXiv:hep-th/0703267].
\bibitem{Isozumi:2004jc}
Y.~Isozumi, M.~Nitta, K.~Ohashi and N.~Sakai,
Phys.\ Rev.\ Lett.\  {\bf 93}, 161601 (2004)
[arXiv:hep-th/0404198].
\bibitem{Eto:2005yh}
M.~Eto, Y.~Isozumi, M.~Nitta, K.~Ohashi and N.~Sakai,
Phys.\ Rev.\ Lett.\  {\bf 96}, 161601 (2006)
[arXiv:hep-th/0511088].
\bibitem{Eto:2006pg}
 M.~Eto, Y.~Isozumi, M.~Nitta, K.~Ohashi and N.~Sakai,
 J.\ Phys.\ A  {\bf 39}, R315 (2006)
 [arXiv:hep-th/0602170].
\bibitem{Eto:2008yi}
M.~Eto, T.~Fujimori, S.~B.~Gudnason, K.~Konishi, M.~Nitta, K.~Ohashi
and W.~Vinci, 
Phys.\ Lett.\  B {\bf 669}, 98 (2008)
[arXiv:0802.1020 [hep-th]].
\bibitem{Eto:2009bg}
M.~Eto {\it et al.},
JHEP {\bf 0906}, 004 (2009)
[arXiv:0903.4471 [hep-th]].
\bibitem{Slansky:1981yr}
R.~Slansky,
Phys.\ Rept.\  {\bf 79}, 1 (1981).
\bibitem{Auzzi:2008wm}
 R.~Auzzi, M.~Eto, S.~B.~Gudnason, K.~Konishi and W.~Vinci,
 Nucl.\ Phys.\  B {\bf 813}, 484 (2009)
 [arXiv:0810.5679 [hep-th]].

\end{thebibliography}
\end{document}